\documentclass[12pt]{article}

\def\Journal#1#2#3#4{{#1} #2 (#4) #3}

\def\NPA{{Nucl. Phys.} A}
\def\NPB{{Nucl. Phys.} B}
\def\PLB{{Phys. Lett.}  B}
\def\PR{{Phys. Rep.}}
\def\PRL{Phys. Rev. Lett.}
\def\PRC{{Phys. Rev.} C}
\def\PRD{{Phys. Rev.} D}

\def\ZPC{{Z. Phys.} C}

\def\P{{\mathbf p}}
\def\Q{{\mathbf q}}
\def\K{{\mathbf k}}

\def\mbf#1{{\mathbf #1}}

\def\a{\alpha}

\def\d{\delta}
\def\e{\epsilon}

\def\p{\pi}
\def\q{\theta}
\def\m{\mu}
\def\n{\nu}

\def\s{\sigma}
\def\t{\tau}
\def\z{\zeta}

\def\cm{{\cal M}}

\def\intps{\int \frac{d^3 \P}{(2\pi)^3}}

\def\exp{\mbox{\rm exp}}

\def\del{\mbox{$\partial$}}
\def\ncdot{\!\cdot\!}

\def\be{\begin{equation}}
\def\ee{\end{equation}}
\def\bea{\begin{eqnarray}}
\def\eea{\end{eqnarray}}
\def\eref#1{Eq.~(\ref{#1})}

\newcommand{\ncom}{\newcommand}
\ncom{\vo}[1]{{\mathbf #1}}
\ncom{\vmo}[1]{{|\mathbf #1|}}
\ncom{\vt}[2]{({\mathbf #1}-{\mathbf #2})}
\ncom{\lan}{\langle}
\ncom{\ran}{\rangle}
\ncom\nonum{\nonumber \\}
\ncom\fx{\!\!\!\!}
\ncom\gsim{\mbox{\raisebox{-0.6ex}{\ $\stackrel {>}{\sim}$\ }}}
\ncom\lsim{\mbox{\raisebox{-0.6ex}{\ $\stackrel {<}{\sim}$\ }}}

\ncom{\half}{{1\over 2}}
\ncom{\third}{{1\over 3}}
\ncom{\fourth}{{1\over 4}}
\ncom{\fifth}{{1\over 5}}
\ncom{\sixth}{{1\over 6}}

\begin{document}

\begin{titlepage}

\begin{centering}
\null
\vspace{1cm}
{\bf THERMAL AND CHEMICAL EQUILIBRATION\\ \ \\IN A GLUON PLASMA}


\vspace{2cm}
\centerline{S.M.H. Wong}

\vspace{1cm}
\em \footnote{Laboratoire associ\'e au Centre National
de la Recherche Scientifique}LPTHE, Universit\'e de Paris XI, 
B\^atiment 211,\\ F-91405 Orsay, France

\end{centering}

\vspace{2cm}
\begin{center}
{\bf Abstract}
\end{center}

\vspace{0.5cm}

We show the evolution of a gluon plasma towards equilibrium
starting at some early moment when the momentum distribution 
in the central region is momentaneously isotropic. Using HIJING 
results for Au+Au collision as initial input, we consider thermalization 
and chemical equilibration simultaneously at both LHC and RHIC 
energies. Thermalization is shown to be driven chiefly by inelastic 
process in our scenario contradicting common assumption that this
is the role of elastic process. We argue that only the inelastic
dominancy depends on the initial conditions but not the 
dominance itself.

\vspace{2.0cm}
\noindent LPTHE-Orsay 96/07

\addtocounter{page}{1}
\thispagestyle{empty}
\pagebreak
\end{titlepage}

\section{Introduction}

In future heavy ion collision experiments at the Relativistic
Heavy Ion Collider (RHIC) at Brookhaven and at the Large 
Hadron Collider (LHC) at CERN, one hopes to create a new state 
of matter, the quark-gluon plasma. In order to clearly 
distinguish this from the possible alternative of a hadron gas, 
one would like to know the state of the plasma in its different
stages throughout its lifetime. With this knowledge, the
conditions under which different possible signatures of
the plasma are produced and the time at which they are produced 
can be compared with those of the hadron gas \cite{Letetal}. 
It is precisely the difference of these conditions and
the time dependence which could permit a possible distinction 
from a hadron gas and a clear identification of a quark-gluon
plasma. There have already been numerous works in the
direction of the possible signatures and also a number
of investigations on the various time scales and stages 
of the plasma. In this work, we will focus on the latter.

In the following, we will not attempt to match the 
comprehensiveness and details of the semi-classical parton 
cascade model \cite{Geig1,Geig2} which models the evolution 
in relativistic heavy ion collisions by tracing the phase space 
history of the particles and combining perturbative QCD with 
relativistic transport theory. Our approach will base on rather
simple assumptions and initial conditions which have already been
made used of in the previous studies of thermalization and 
parton chemical equilibration \cite{Biro,Levai,Wang,Heis}. 

In many previous studies 
\cite{Letetal,Biro,Levai,Wang,Shur,Alth}
of relativistic heavy ion collisions, a convenient assumption and 
starting point is kinetic equilibration is very rapid $\leq 1$ fm
and hydrodynamic expansion is well underway. As has been pointed
out in \cite{Heis,Dan}, collisions are not sufficiently rapid
at early times to maintain thermal equilibrium in an expanding
plasma, so the system is more likely to be in a situation between
free streaming, where there is no collision or the collision
effects are completely negligible, and hydrodynamic expansion when
local kinetic equilibrium has been achieved. Various studies
have been made in determining the relation between collision time
$\theta$ and the degree of equilibration of the plasma 
\cite{Heis,Baym,Gavin} based on relaxation time approximation and 
assumed simple power behaviour of $\theta \sim \t^p$. In this work, we 
will also look at the collision time and thermalization but we do not 
assume a priori any form for $\theta$, rather it will be determined 
entirely from the interactions. Our starting point will be different
from many previous works in that we start somewhere between free 
streaming and hydrodynamic expansion. Some initial steps in this 
direction have been taken by Heiselberg and Wang in \cite{Heis} but 
they only considered elastic interactions and only in the time range 
$\t_0 \lsim \t \ll \q$, where $\t_0$ is the initial time which will
be further explained below. Following their initial line of approach, 
we attempt to give a more complete treatment. We will do this under 
certain assumptions and approximations which will be explained 
in the subsequent sections. The method that will be shown here
allows the treatment of the out of equilibrium plasma with
comparative ease. We will illustrate this approach by applying
it to the simpler case of a pure gluon plasma. Thermal as well as 
chemical equilibration of the gluon plasma will be investigated. 
That of a quark-gluon plasma will be dealt with in a subsequent 
paper \cite{Wong}.

For the initial conditions, we will use results obtained from 
the Monte Carlo HIJING model \cite{Gyu1,Gyu2,Gyu3} as have been used
in \cite{Biro} and later improved in \cite{Levai,Wang} but we will
interpret them in a slightly different way so that our initial
temperatures $T_0$ and fugacities $\l_0$ are not the same as theirs
although we use the same initial gluon energy densities $\e_0$ 
and number densities $n_0$.  

In principle, one could vary the initial conditions and look at the
resulting evolution of the plasma as have been done in \cite{Shur}
since HIJING predictions for the energy densities and number densities
are in the low end in comparison with, for example, those from the
parton cascade model \cite{Geig2} but we will not do that here. 
We would rather concentrate on one set of specific initial conditions 
each at RHIC and LHC energies and look at the thermalization,
isotropy of momentum distribution and the time dependence of fugacity, 
collision time $\q$ and a modified form of the non-equilibrium colour 
screening length. We show, in particular, inelastic process dominates
over the elastic one in the approach towards equilibrium and 
this is independent of initial conditions. The degree of this
dominance, however, is dependent on initial conditions. 

The paper is organized as follows. In Sect. \ref{sec:Boltz}, the various
assumptions, approximations and approach will be described. 
Sect. \ref{sec:rates} is devoted to calculating rates for the
interactions that we will consider and the set of equations to use for
chemical equilibration will be given. Initial conditions used and the 
method of computation will be outlined in Sect. \ref{sec:I.C.}. 
Lastly, the results of the evolution of the gluon plasma will be 
shown and discussed in Sect. \ref{sec:equil}.

\section{Boltzmann Equation and Relaxation Time Approximation}
\label{sec:Boltz}

In a many-body system, it is expedient to use particle distributions
$f(\mbf p,\mbf r,\t)$ as well as collective variables 
energy density $\e$, number density $n$, entropy density $s$ etc.
or at least local collective variables to describe the spatial as
well as temporal variation of the system. This is what we will
do here but we concentrate mainly in the central collision region
which we assume to be essentially uniform in space. 
Ideally, one would like to describe $f(\mbf p,\t)$ by a set of fully
relativistic and quantum transport equations derived from first
principle of QCD \cite{Winter,Heinz1} since a quark-gluon plasma is 
inherently a quantum system. But unfortunately, this has not yet 
been realized although various advances have been made in that 
direction based on constructing Wigner operators from the fields 
and deriving their equations of motion from the field equations 
\cite{Elze1,Heinz2}. The difficulty is partly due to the
complexity of the equations of motion for the Wigner operators
which requires, in the end, one to make a semi-classical expansion 
in powers of $\hbar$ and partly due to the gauge degrees of freedom 
which make it hard to identify a gluon distribution operator 
\cite{Elze2}. These gauge degrees of freedom of QCD have proved to 
be a stumbling block in identifying a Wigner operator as the gluon 
distribution operators. Recent works \cite{Heinz3,Heinz4} suggest that 
this latter problem could be overcome by choosing the radial gauge 
but one has yet to derive the collision terms.

Under these circumstances, one has to make do with relativistic
but semi-classical approach such as using Boltzmann equation
\be \Big ({\del \over {\del t}} +\mbf v \ncdot 
    {\del \over {\del \mbf r}} \Big )f(\mbf p,\mbf r,t)
    =C(\mbf p,\mbf r,t)  \; ,
\label{eq:boltzeq}
\ee
and combining it with collision term $C$ obtained from QCD. 
Assuming one-dimensional expansion along the z-axis \cite{Bjork} 
and boost invariance or rapidity independence of the 
central region, Baym \cite{Baym} showed that \eref{eq:boltzeq}
can be rewritten as 
\be \Big ({{\del f(p_\perp, p_z,\t)} \over {\del \t}} \Big )
    \Big |_{p_z \t}=C(p_\perp,p_z,\t)  \; .
\label{eq:baymeq}
\ee

As in \cite{Heis,Baym}, we use the relaxation time approximation 
for the collision term
\be C(\mbf p,\t)=-{{f(\mbf p,\t)-f_{eq}(\mbf p,\t)} \over \q(\t)}
\label{eq:Crlx}
\ee
which enables us to write down a solution to the Boltzmann 
equation \eref{eq:baymeq}
\be f=f_0(p_\perp,p_z \t/\t_0) e^{-x}
      +\int^x_0 dx' e^{x'-x} 
      f_{eq}(\sqrt{p^2_\perp+(p_z \t/\t')^2},T_{eq}(\t')) \; ,
\label{eq:baymeqsol}
\ee
where 
\be x(\t)=\int^\t_{\t_0} d\t'/\q(\t') \; ,
\label{eq:x}
\ee
given some initial distribution $f_0$ at $\t_0$. $f_0(p,\t)$ is
the free streaming, $C=0$, solution to \eref{eq:baymeq}.
Our solution differs from that in \cite{Heis}. In their case,
they considered only elastic interactions and so had to keep a
chemical potential $\m(\t')$ in $f_{eq}$. We will not restrict 
ourselves to only these interactions. On the contrary, we will
use the interaction $gg\longleftrightarrow ggg$ to determine
$\q(\t)$. $T_{eq}(\t)$ in $f_{eq}$ is determined from energy
conservation
\be \e(\t)=\e_{eq}(\t)=a_2 T^4_{eq}(\t) \; ,
\ee
with $a_2=8\p^2/15$. It is the momentaneous target temperature
that the gluon plasma is striving to reach. 
The associated energy conservation equation is \cite{Baym}
\be {{d \e} \over {d \t}}+{{\e+p_L} \over \t} =0 \; ,
\ee
and $p_L$ is the longitudinal pressure in the z-direction given by 
\be p_L(\t)=\n \intps {{p_z^2} \over p} f(p_\perp,p_z,\t) \; ,
\label{eq:presl}
\ee
with $\n=2\times 8=16$ for the multiplicity of gluons.
Along the same line, we can also defined $p_T$, the transverse
pressure by replacing $p_z$ by $p_x$ or $p_y$ in \eref{eq:presl}.
These pressures will be used to check the degree of anisotropy
of momentum distribution in the central region of the plasma.

\section{Rate Equations}
\label{sec:rates}

To consider chemical equilibration, we require the rate equations
for gluon creation or destruction. From \cite{Mat} and 
\eref{eq:baymeq}, we have the following rate equation for the
gluon number density
\be {{\del n} \over {\del \t}}+{n \over \t}
    =\s \intps C(\mbf p,\t) = R(\t) \; ,
\label{eq:rateq1}
\ee
and using \eref{eq:baymeq} and \eref{eq:Crlx}, we also have
\be R(\t)={{n_{eq}(\t)-n(\t)} \over {\q(\t)}} \; ,
\label{eq:rateq2}
\ee
where $n_{eq}(\t)=a_1 T^3_{eq}(\t)$ and $a_1=16 \z(3)/\p^2$.
The $\s$ in \eref{eq:rateq1} is a numerical factor to take into
account of multiplicity and symmetry of the permutations of the
gluons in $C$. The two equations \eref{eq:rateq1} and 
\eref{eq:rateq2} allow us to determined $\q(\t)$ at any instance 
in time. 

For the r.h.s. of \eref{eq:rateq1}, we take this to be the
simplest gluon multiplication process
$gg \longleftrightarrow ggg$
\bea \s \intps C(\mbf p,\t) \fx &=& \fx \fourth (2\p)^4 
     \n^2 \prod^5_{i=1} {{d^3 \P_i} \over {(2\p)^3 2 p_i}} 
     |\cm_{gg\rightarrow ggg}|^2 \d^4(p_1+p_2-p_3-p_4-p_5) \nonum
     \fx & & \fx \times
    [f_1 f_2(1+f_3)(1+f_4)(1+f_5)-f_3 f_4 f_5(1+f_1)(1+f_2)] \nonum
     \fx & & \fx \times \q (\Lambda-\t_{QCD}) \; .
\label{eq:g_no_source}
\eea
The various parts of this formula need some explanation, we do this
one by one. Here we only consider small angle scattering and without 
loss of generality $p_5$ is taken to be a soft gluon radiation around
zero rapidity in the centre of momentum frame so that
we can use Bertsch and Gunion formula \cite{Gunion} for the
amplitude. In that frame of the two incident gluons, where they 
approach each other on the z-axis, this is
\be |\cm_{gg\rightarrow ggg}|^2= \Big ({{4 g^4 N^2} \over {N^2-1}}
    {s^2 \over {(\mbf q_\perp^2+m_D^2)^2}} \Big )
    \Big ( {{4 g^2 N \mbf q_\perp^2}\over 
    {\mbf k_\perp^2 [(\mbf k_\perp-\mbf q_\perp)^2+m_D^2]}} \Big )
\label{eq:gun&bert}
\ee
where $\K_\perp$ and $\Q_\perp$ are the perpendicular component of $\P_5$ 
and that of the momentum transfer in that frame respectively. $N=3$ 
is for SU(3) of QCD and $s$ is the total centre of mass energy. 
The first part on the r.h.s. within round brackets is the squared
modulus of the small angle elastic scattering amplitude for 
$gg\rightarrow gg$ and the second part is essentially the soft 
gluon emission spectrum. Since the centre of momentum frame of any
two incident gluons are very unlikely to be the same as the plasma 
rest frame, we need a general formula for this amplitude. 
In the plasma rest frame, one simply substitues 
\bea (\K_\perp-\Q_\perp)^2 \fx &=&\fx 4 \, p_1 \ncdot p_3 \;
     p_2 \ncdot p_3/s \\
    \Q_\perp^2 \fx &=&\fx 4 \, p_1 \ncdot p_4 \; p_2 \ncdot p_4/s \\
    \K_\perp^2 \fx &=&\fx 4 \, p_1 \ncdot p_5 \; p_2 \ncdot p_5/s
\eea
in \eref{eq:gun&bert}. 

We have infrared regularized the squared modulus of the amplitude 
\eref{eq:gun&bert} with the Debye screening mass $m_D^2$ 
which is time dependent in the present case. We take it to
be defined at leading order in $\a_s$ by
\be m_D^2(\t)= -8\p\a_s N\intps {{\del f(\P,\t)}\over {\del \vmo p}} 
    \; .
\ee
Note that the screening mass should be directional dependent in 
general \cite{Biro2,Eskola1} but rather than keeping tracks of the 
directions of the gluons, we simplified this by removing the directional
dependence. The present form can be thought of as an averaged
screening mass between the transverse and longitudinal screening.
We will explain this in some details elsewhere \cite{Wong}.
Here $f$ is not of the Bose-Einstein form when the plasma is
out of equilibrium but is given by \eref{eq:baymeqsol}. It reduces, 
of course, to the familiar $m_D^2=4 \p \a_s T^2$ when in 
equilibrium. This provides us with a general Debye screening
mass to screen off infrared divergences even in out of 
equilibrium situation which is the subject of this paper. 

As in \cite{Biro,Levai,Wang}, we use the approach of Gyulassy and 
Wang \cite{Gyu4,Gyu5} to deal with the Landau-Pomeranchuk-Migdal
(LPM) suppression of soft gluon radiations due to multiple 
scattering of the parent gluon in a medium. This is taken
into account in the form of the $\q$-function in 
\eref{eq:g_no_source}. The basis of this approach is to 
include only the non-suppressed Bethe-Heitler limit but
the suppressed emission between that and the factorization limit 
is completely dropped. Ideally, one would also like to include
these emissions. The $\q$-function requires that the 
mean free path, $\Lambda$, of the gluon which we take from the 
small angle $gg$ elastic scattering cross-section \cite{Biro,Levai,Wang},
\be {{d \s^{gg\rightarrow gg}_{el}} \over {d \Q^2_\perp}}
    ={9 \over 4} {{2 \p \a_s^2} \over {(\Q^2_\perp+m^2_D)^2}}
\ee
\be \Lambda^{-1} = n \int^{s/4}_0 d \Q^2_\perp 
    {{d \s^{gg\rightarrow gg}_{el}} \over {d \Q^2_\perp}}
    = {{9\p \a_s^2 s n}\over {8 m^2_D (s/4+m_D^2)}}
\ee
to be larger than the formation time $\t_{QCD}=k_0/\K_\perp^2$ of 
the soft gluon. This LPM effect on the radiative energy loss of a 
fast energetic parton previously derived in \cite{Gyu5} has been 
improved and rederived recently in \cite{Baier} by including 
interference graphs where the emitted gluon also undergoes 
multiple scattering with the scattering centres. 
However, the derived soft radiation LPM spectrum has no
dependence on the direction of the emitted gluon. This has to be
integrated out for the radiation density to make sense. 
As such, their result cannot be used in the present problem.
One will have to be contented to include only the Bethe-Heitler
non-suppressed emission therefore we use the same $\q$ function
as in \cite{Biro,Levai,Wang}. 

At this point, a remark is in order concerning large angle
scattering non-soft gluon emission. In \cite{Shur}, Shuryak and 
Xiong found that large angle scattering is also important and higher
order tree diagrams of maximal helicity violating amplitudes,
the so called Kunszt-Stirling \cite{Kunszt} corrected Park-Taylor 
formula \cite{Parke} for gluon multiplication are not negligible. 
In performing their calculation, LPM suppression has been totally
neglected. As far as the author is aware, possible interference
effect on non-soft gluon emission in multiple scattering has not
been worked out. May be this is negligible for non-soft radiation
if one is permitted to generalize the language of the soft emission
to the non-soft case. That is by arguing that the gluon mean free path
is approximately the same or larger than the coherence length of
these radiations, but the latter is not defined for non-soft
radiation. Facing the choice of giving up LPM entirely or 
including large angle scatterings, we choose the first alternative.
The latter will be investigated in a future work where some form
of LPM or more general interference effect due to multiple scattering
will have to be included.

\section{Initial Conditions for the Computations}
\label{sec:I.C.}

As in \cite{Biro,Levai,Wang,Heis}, we consider the following
scenario. Initially, the two incoming heavy ions which have
been accelerated to very high energies, 200 GeV/nucleon at RHIC 
and up to 6.3 TeV/nucleon at LHC, along the z-axis collide, 
penetrate each other leaving behind in the central region a gas of
secondary partons produced via minijets \cite{Eskola2}. Since 
most energies are in the longitudinal direction, the momentum 
distribution is mainly along the beam direction. As mentioned in 
the introduction, collision time is larger than the expansion 
time so initially the partons are more likely to stream freely in
the longitudinal direction gradually suppressing the momentum 
distribution in that direction in the central region. One can
understand this by looking at the form of $f_0$ in \eref{eq:fzero}.
At certain moment $\t_0$, longitudinal momentum distribution
will be reduced to such an extent that it becomes the same 
as the transverse momentum distribution, or in other words,
the momentum distribution becomes isotropic. Assuming a
thermal distribution at this time $\t_0$ fixes $f_0(\P,\t_0)$.
The initial temperature $T_0$ and fugacity $l_0$ can be determined 
from the gluon energy density and number density. These values will 
be taken from the improved version \cite{Levai} of \cite{Biro}
but we obtain the temperature and fugacity in a slightly
different way. In \cite{Biro,Levai}, they used the fugacity
factorized thermal equilibrium form of $f$, i.e. 
\be f(\P,l)={l \over {\exp (\vmo p/T)-1}}
\ee
for all values of $l$. As was pointed out in their work, this is a 
good approximation only when $l\approx 1$. Here we keep the original 
form of $f$ at least whenever it is in thermal equilibrium. We get 
the initial temperature $T_0$ and fugacity $l_0$ at $\t_0$ by 
observing that $l_0$ is small initially, at least in HIJING results, 
when we can approximate $f$ by the Boltzmann form $f=l \; e^{-p/T}$ 
so the energy density $\e_0$ and number density $n_0$ are given by
\be \e_0 = 3 \,\n\, l_0\, T^4_0/\p^2 \mbox{\hskip 1cm and \hskip 1cm} 
     n_0 = \n\, l_0\, T^3_0/\p^2 \; ,
\ee
respectively. The extracted $T_0$ and $l_0$ for Au+Au collision at 
LHC and at RHIC are shown in Table 1. Knowing 
\be f_0(\P,\t_0)={l_0 \over {\exp (\vmo p/T_0)-l_0}} \; ,
\ee
we can deduce the time dependent form for it, since it is the free 
streaming solution to \eref{eq:baymeq} \cite{Heis,Baym}
\be f_0(p_\perp,p_z,\t)={l_0 \over
    {\exp (\sqrt {p^2_\perp+p_z^2 \t^2/\t^2_0}/T_0)-l_0}} \; .
\label{eq:fzero}
\ee
Using this, we can calculate the rate $gg\longleftrightarrow ggg$ 
at $\t_0$ and hence $\q_0$. With these initial values, we start the
evolution of the plasma.

The evolution of the gluon plasma is performed as follows. Phase 
space integrals for calculating the gluon multiplication rate are
performed by an iterative and adaptive Monte Carlo method 
VEGAS \cite{Lepage}. All time integrals are calculated by discretizing 
time $\t$ so that any function $G(\t)$ depending on a time integral 
of another function $g(\t,\t')$ is re-expressed as
\be G(\t)=\int^\t_{\t_0} d\t' g(\t, \t') \;
     \Longrightarrow \; G(\t_n) \approx
     \sum^{n-1}_{i=0} \triangle \t \; g(\t_n,\t_i)  \; ,
\ee
where $\t_i=\t_0+i \triangle \t$ and $\triangle \t$ is some small 
chosen time step. Then $G(\t_n)$ is determined entirely by variables 
of the preceding time steps from $\t_0$ to $\t_{n-1}$.
Given all the variables at $t_0$, the same set of variables at
$\t_1$ can be determined. This procedure is repeated for obtaining
the variables at $\t_2$ from those at $\t_0$ and $\t_1$ and so on.
The set of variables to determine are $\e(\t)$ for obtaining
$T_{eq}(\t)$ which enables the determination of $m_D^2(\t)$. This
modified screening mass is required for calculating $R(\t)$ which
together with $n(\t)$ and $T_{eq}(\t)$ allow us to obtain $\q(\t)$.
Both $\q(\t)$ and $T_{eq}(\t)$ are essential for finding $f(\P,\t)$
which is itself needed in $R(\t)$ and for the calculations of
the various collective variables.

\section{Equilibration of the Gluon Plasma}
\label{sec:equil} 

The results of the evolution towards equilibrium of a gluon plasma
for two sets of initial conditions for Au+Au collisions at LHC and 
at RHIC are presented in Fig. 1, 2 (LHC) and Fig. 3, 4 (RHIC). These
are results calculated using $\a_s=0.3$ and $\triangle \t=0.05$ fm/c. 
Using smaller values of $\triangle \t$, for example, $\triangle \t=0.02$ 
fm/c shifts the various values by amounts ranging up to 6.25\%. 
Since we are not going after high numerical accuracy, we will settle 
for the present value. In any case, there is no qualitative change by
varying $\triangle \t$. Unless explicitly stated otherwise, the dashed 
curves in most figures are the equilibrium values associated with 
$f_{eq}$ in \eref{eq:Crlx}. The solid curves are the actual results. 
One sees that the two curves approach each other with increasing 
time. How close the solid and dashed curves are is a measure of how 
close the plasma is to equilibrium. The computation is stopped when 
the estimated temperature falls below 200 MeV. This and the fugacity
are estimated by using the expression for the near full equilibrium 
energy and number density
\be \e = a_2 \, l \, T^4 \mbox{\hskip 1cm \rm and \hskip 1cm}
     n = a_1 \, l \, T^3
\label{eq:near_eq_en}
\ee
which are valid when $l$ is close to one, so that
\be T_{est} = a_1 \,\e/ (a_2 \, n) \mbox{\hskip 1cm \rm and \hskip 1cm}
    l_{est} = a_2^3 \, n^4/(a_1^4 \, \e^3)  \; .
\ee
Since we have used energy conservation $\e(\t)=\e_{eq}(\t)$ at 
all time so in Fig. 1 and 3 (a) where we show the evolution of the
energy densities, the solid and dashed curves coincide.
Comparing gluon densities in Fig. 1 and 3 (b), the estimated
temperatures in Fig. 1 and 3 (c), the entropy densities in Fig. 1 and
3 (e) and the squared of the modified 
screening mass in Fig. 2 and 4 (b), it is apparent that at LHC, 
the solid and dashed curves are much closer to each other at the 
end of the calculation than they are at RHIC. Those at LHC are 
actually lying on top of each other or at least very close.
The estimated fugacity $l_{est}$ at the end of the calculation is 
about 0.94 at LHC and 0.40 at RHIC. The time evolution of these and 
the estimated temperatures are shown in Fig. 1 and 3 (c). The 
dot-dashed curves are the fugacity evolutions. Note that these
estimated values can only be viewed as indicators. It will be seen
that the plasma have not quite completely thermalized at the end 
of the calculations. 

The entropy density evolutions $s(\t)$ shown in Fig. 1 (e) for LHC
and in Fig. 3 (e) for RHIC are calculated from \cite{Groot}
\be s(\t)=-\n \intps \Big \{f(\P,\t) \ln f(\P,\t)
          -(1+f(\P,\t)) \ln (1+f(\P,\t)) \Big \} \; .
\ee
The dashed lines are $s=4\e/3 T_{eq}$, the ideal gas form of 
the equilibrium entropy. To look at the actual time dependence of 
the generated entropy, we plotted the product of the entropy 
density and $\t$, $s \t$, which is proportional to the total 
entropy in the central region in Fig. 1 and 3 (f). At LHC, the 
entropy no longer increases or stopped being produced at the end at 
$\t=$ 8.75 fm/c, whereas at RHIC, it has not yet stopped to increase 
at $\t=$ 3.55 fm/c when the estimated temperature falls below 200 MeV. 

To check for isotropy in momentum distribution, we plotted 
the ratio of longitudinal to transverse pressure in Fig. 1 
and 3 (d). These pressures are as defined in Sect. \ref{sec:Boltz}. 
The solid curve in each case is this ratio while the dashed curve 
is the ratio of a third of the energy density to the transverse 
pressure for comparison. The plasma starts off in an isotropic 
momentum configuration and becomes anisotropic due to the 
tendency of the plasma to free stream in the z-direction. This
anisotropy attains its maximum at about 1.35 fm/c at LHC 
and 2.2 fm/c at RHIC. Since the plasma is approaching 
equilibrium, both $p_L/p_T$ and $\e/3 p_T$ must approach 
1.0 with increasing time. The figures show the plasma has 
not quite regained isotropy in momentum distribution 
although it is much closer at LHC than at RHIC. 
Note that $p_L <\e/3$ except at the start therefore 
less work is required for the expansion than in the 
case of hydrodynamic expansion when $p_L=\e/3$. The 
consequence is the energy density drops less fast with increasing
$\t$. Note that this does not necessarily mean that the plasma can last
longer than if it underwent hydrodynamic expansion. Particle
production rate must also be compared because although this does not
change the energy density, it does reduce the temperature whenever
the concept of a temperature makes sense. If the rate is not 
so rapid that it over compensates for the reduction in $p_L$,
then indeed the plasma can last longer than that undergoing hydrodynamic 
expansion before phase transition takes place.

To look at the relative importance of free streaming and 
collisions, we plotted in Fig. 2 and 4 (a), the modified screening
mass squared (solid curves) and that of free streaming when
collisions have been switched off. The modified screening mass 
squared is
\be m_D^2(\t)={\t_0 \over \t} j(\t_0/\t) \, m_{D\, 0}^2 \, e^{-x(\t)}
       +4 \p \a_s e^{-x(\t)} \int^\t_{\t_0} {{d\t'} \over {\q(\t')}}
        {\t' \over \t} j(\t'/\t) \, T^2_{eq}(\t') \, e^{x(\t')}
\ee
where 
\be j(r) = {{\sin^{-1} \sqrt {1-r^2}} \over {\sqrt {1-r^2}}}
\ee
and that of free streaming is obtained by 
letting $\q \rightarrow \infty$. The latter is a decreasing function
of time so free streaming increases the screening length inside
the plasma since the gluon density decreases with $\t^{-1}$.
Whereas collisions tend to shorten the screening length or increase
the screening mass due to the equilibrium value is several times higher.
We see that even at the beginning, collision effect is already very 
strong especially at LHC where $\q_0 \sim 0.84$ when $\t_0 \sim 0.5$
so the plasma starts very much in the mixed phase of free streaming
and hydrodynamic expansion. This is less so at RHIC with 
$\q_0 \sim 2.7$ when $\t_0 \sim 0.7$ but the collision effect is
still strong. In both cases, the screening mass increases first with
time (more rapid increase at LHC than at RHIC) before it decreases
eventually because the plasma is undergoing expansion and therefore
cooling. Fig. 2 and 4 (b) show the convergence of $m_D^2(\t)$ with
the equilibrium value with increasing $\t$. Again, we see a better
convergence at LHC than at RHIC.

Another evidence for the strong collision effects is the time
dependence of the collision time $\q(\t)$ plotted in 
Fig. 2 and 4 (c) (These curves are relatively rough but because of
the time required for the computations, we do not attempt to get
smoother curves. In any case, how $\q$ is evolving with time
is clear.). One expects free streaming to drive the system
further out of equilibrium and hence increases $\q$. Fig. (c)'s 
show quite the opposite behaviour of a clear initial drop in $\q$
in each case. A much steeper drop at RHIC than at LHC is seen. 
These initial drops are followed by relatively slow increase
with respect to $\t$. These behaviours are essential for the plasma
to be able to reach equilibrium i.e. they allow $\t$ to catch up
with $\q$ and overtake it. This overtaking of $\q$ by $\t$ is
much more dramatic at RHIC since $\q_0$ is about 4 times larger
than $\t_0$ whereas at LHC, $\q_0/\t_0$ is only about 1.6. The
final slow increase in $\q$ is expected from previous 
calculations of relaxation time \cite{Baym2,Heis2}. They found 
in general when the system is near equilibrium a dependence 
$\t_{relax} \sim 1/T$ on the temperature which means a time
dependence of $\t_{relax} \sim \t^{1/3}$. 
In our case, we can parametrize the rate $R(\t)$ near full 
equilibrium by
\be R(\t) \approx (1-l) \, c \, T^4 = (1-l) \, c' \, a_1 T^4
\label{eq:r_neareq}
\ee
where $c$ and $c'$ are some constants. Using \eref{eq:near_eq_en}, we
get 
\be \triangle n=n_{eq}-n=a_1 \, T^3 \, (l^{3/4}-l) \; .
\ee
From the above two equations, one obtains the near equilibrium
collision time $\q_{l \rightarrow 1}(\t)$
\be \q_{l \rightarrow 1}(\t)={l^{3/4}\over {c' T(1+l^{1/2}) (1+l^{1/4})}}
    \; .
\ee
We have plotted this in Fig. 2 (c) (dashed curve) using $T_{est}$ and
$l_{est}$ but not in Fig. 4 (c) since at RHIC, the plasma is still not 
yet near equilibrium. The constant $c'$ determined from the rate is
roughly $c'\approx 0.16377$. One sees that the solid and dashed curve 
converges at large time. One could also fit a $\t^{1/3}$ curve on the 
same plot (not shown for clarity) which would show $\q$ does have the 
right behaviour for a equilibrated or nearly equilibrated plasma.

Although our $\q$ has a more complicated time dependence than simple
power behaviour which was heuristically assumed in previous works
\cite{Heis,Baym,Gavin} to enable simple analysis, it does satisfy the
generalized condition of Heiselberg and Wang \cite{Heis} for 
equilibration at all time. Their analysis for $\q \sim \t^p$ 
showed thermalization would be achieved for $p<1$ otherwise $\q$
would increase faster than $\t$ or collision would always be slower 
than the expansion. For general $\q$, this can be reworded in terms 
of $x(\t)$ defined in \eref{eq:x} as $x \rightarrow \infty$ as 
$\t \rightarrow \infty$ for thermalization, if 
$x \rightarrow x_{max}< \infty$ as $\t \rightarrow \infty$, then 
thermalization cannot be realized.

Finally, a word must be said about elastic collisions
$gg\rightarrow gg$ which have almost been left out entirely
since they drop out of \eref{eq:rateq1}
(their only role so far is in determining the cross-section).
In \cite{Heis}, by considering only elastic processes, $\q$ has been 
determined from the collision entropy density rate
\be {{\del s} \over {\del \t}}\Big |_{coll}
    = -\s' \intps C(\P,\t) \ln \Big ({f \over {1+f}}\Big )
\label{eq:srate}
\ee
where $\s'$ is again some numerical factor for multiplicity etc..
They found that $\q$ had only logarithmic time dependence.
Their result, which is approximately valid in the range 
$2\t_0 \lsim \t \ll \q$ and after being readjusted to our initial 
conditions, is
\be {1 \over \q} \simeq 1.55 \, \a_s^2 \, l_0 \, T_0
    \ln \Big ({{3 \p \t}\over {2 \, l_0 \, \a_s \, \t_0}} \Big )
    \ln \Big ({{2\t} \over \t_0} \Big )  \; .
\label{eq:estheta}
\ee
Using $\t$ between $2 \t_0 \lsim \t \lsim \q/2$, we find $\q$ varies 
between $4.28$--$3.53$ fm/c from $\t \simeq 1.40$ fm/c to 
$\t \simeq 1.76$ fm/c at RHIC, which is about 2.0 times higher than 
our $\q$ in the same time range. At LHC, the above time range does 
not exist and so \eref{eq:estheta} cannot be applied. This shows that 
inelastic process $gg\longleftrightarrow ggg$ seems to be more efficient
for thermalization than elastic $gg\rightarrow gg$ at least at RHIC. 
To verify this, we have checked using \eref{eq:srate} how important 
is the entropy generated by elastic scattering when compared to 
that by inelastic scattering both at RHIC and at LHC. The 
result is small. We have to conclude that in our scenario 
the latter is the chief driving force for thermalization
which contradicts the common assumption that thermalization
is mainly the role of elastic processes. This conclusion is perhaps 
not too surprising since we have started in a momentaneously 
thermalized plasma with some small initial fugacities when all 
elastic processes are ``out of action''. Our figures show that the
plasma quickly reaches the point of greatest departure from
equilibrium before heading back progressively. In order for 
elastic scatterings to play a more prominent role, there need 
to be some means for an explosive rise in rate before that 
point\footnote{We can reasonably expect the elastic
scattering rate maximizes at this point. This is within the
time range in which we compared $\q$'s above.} has been 
reached. This means is, however, missing. It must be said that
although our conclusion on the dominance of inelastic over
elastic process is independent of the initial conditions,
the degree of dominance is not. Had we started with some 
much larger initial fugacities, say $l_0 \sim 1$, then 
the initial gluon multiplication rate $R(\t=\t_0)$ would 
be greatly reduced and this will affect the dominancy. 
This can be understood in the following way. Since 
one cannot perturb an equilibrated gluon plasma 
kinetically without doing so chemically at the same time 
\footnote{The reverse is, however, not true. One can see 
this by finding a small perturbation that will hold chemical 
but not kinetic equilibrium.}, any perturbation will 
always drive a plasma in equilibrium further off chemical 
than kinetic equilibrium. In our case, the perturbation
is the one-dimensional expansion and it is not obvious that
the plasma has been chemically perturbed unlike the perturbation 
in the kinetic direction. This is so because the chemical 
perturbation due to the expansion is too small when compared 
to that with which we imposed on the system by having small 
$l_0$'s. For $l_0 \sim 1$, the rate in Fig. 2 and 4 (d) will
not be monotonically decreasing with increasing $\t$. They will 
instead first rise to a peak before coming back down again 
towards zero. A typical sign that the system is going  
out of equilibrium before making the return \cite{Wong}. In 
this case, clear chemical along with kinetic equilibrium 
are further upset by the expansion. Numerically, this
has been verified with $l_0 \sim 0.8$. In this case, the 
relative importance of the two types of process is reduced 
but inelastic remains the more important one.
To summarize, it is because of the fact that chemical 
equilibrium is always further away than kinetic equilibrium, 
inelastic process will always be significant in relation to
elastic process in equilibration. A small $l_0$ will give it 
a bigger advantage or increased importance than a $l_0 \sim 1$

\section*{Acknowledgements}

The author would like to thank R. Baier for clearing up delicate 
matters on general interference effect, U. Heinz for useful comments,  
S. Peign\'e and D. Schiff for explanation of LPM. The author
acknowledges financial support from the Leverhulme Trust.

\break
\section*{Figure Captions}

\begin{itemize}

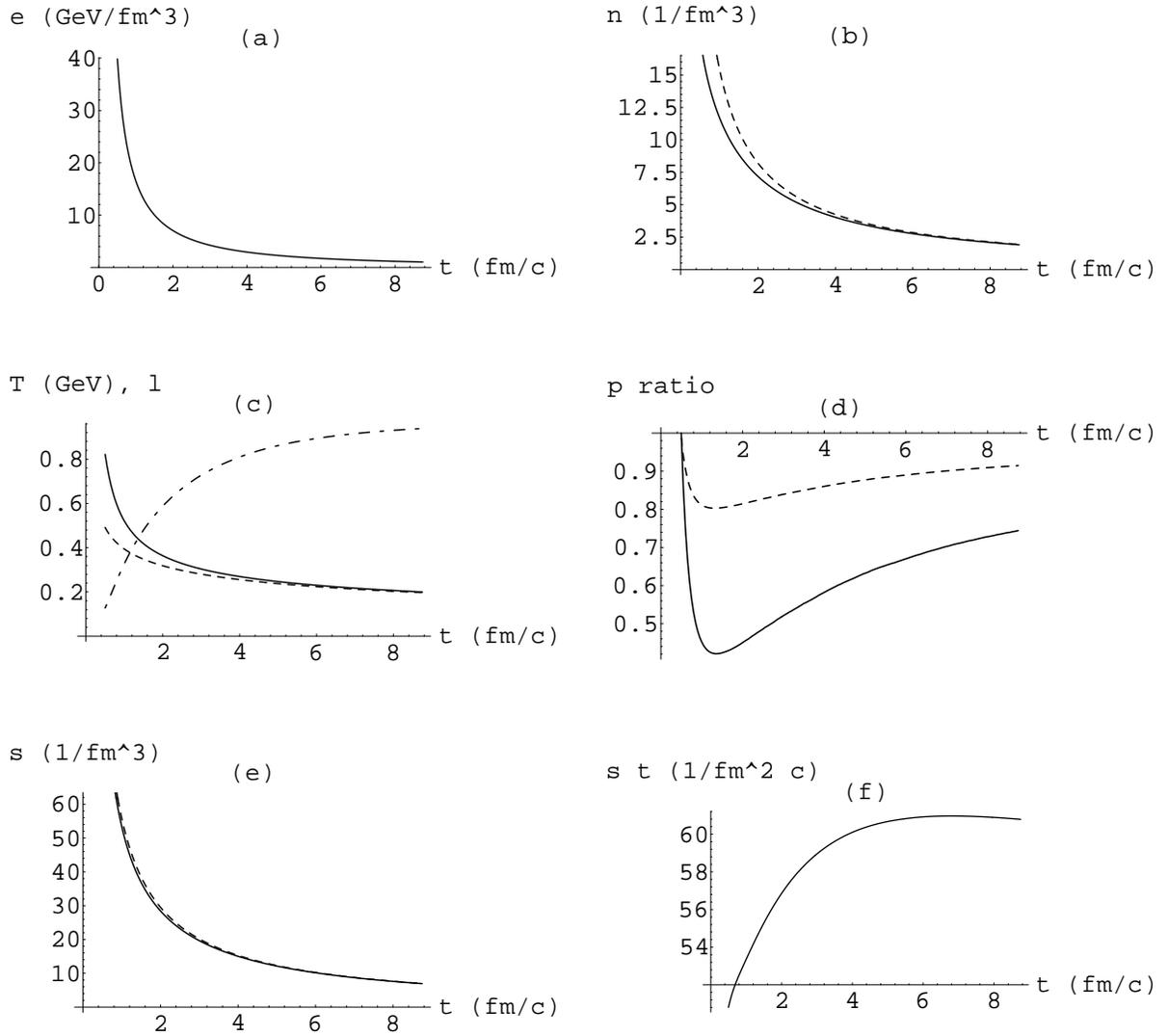
\item[Fig. 1] Results of the evolution of a gluon plasma towards 
equilibrium at LHC energies. The figures are (a) energy density 
$\e$, (b) number density $n$ (solid curve), (c) estimated 
temperature $T_{est}$ (solid curve) and fugacity $l_{est}$ 
(dot-dashed curve), (d) the ratio of longitudinal to 
transverse pressure $p_L/p_T$ (solid curve) and that of
a third of the energy density to the transverse pressure 
$\e/3 p_T$ (dashed curve), (e) entropy density $s$, 
(f) the product of entropy density and time $s \t$. The 
dashed curves in (b), in (c) and in (e) are the equilibrium 
number density, temperature and the ideal gas entropy 
$s=4\e/3 T_{eq}$ associated with $f_{eq}$ respectively. 

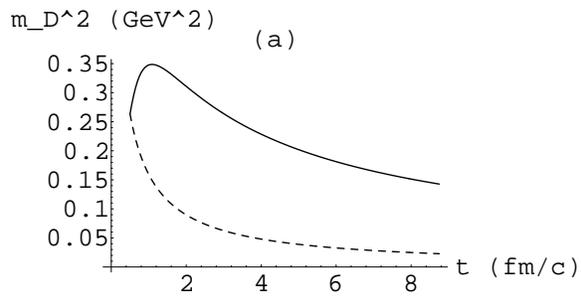
\item[Fig. 2] Results of the evolution of a gluon plasma towards equilibrium 
at LHC energies. The figures are (a) the modified screening mass squared
$m_D^2$ (solid curve) compared with the free streaming case (dashed
curve), (b) $m_D^2$ (solid curve) together with that in equilibrium 
(dashed curve), (c) collision time $\q$ and (d) the rate $R$ for 
$gg\longleftrightarrow ggg$. In this case only (not in Fig. 4),
$\q_{l\rightarrow 1}$ is also shown in (c) (dashed curve).

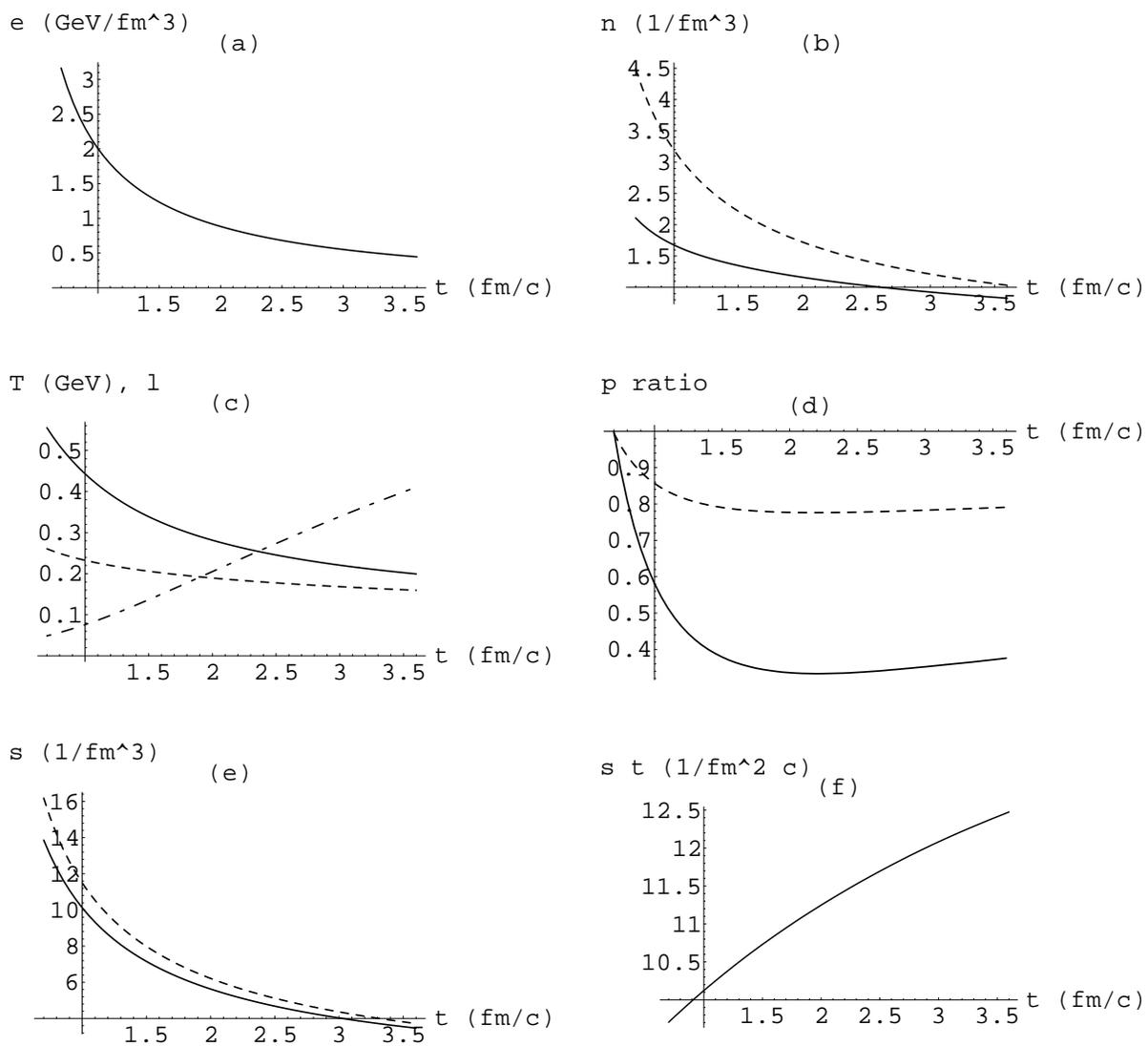
\item[Fig. 3] Results of the evolution of a gluon plasma towards equilibrium 
at RHIC energies. The figures and curves are organized in the same way
as those in Fig. 1.

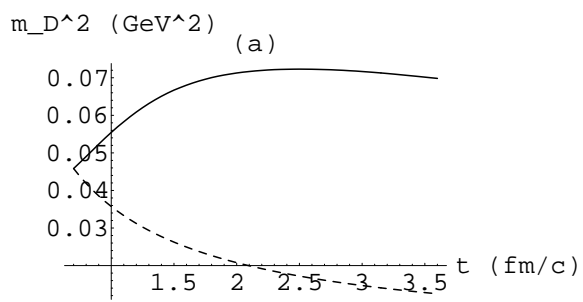
\item[Fig. 4] Results of the evolution of a gluon plasma towards equilibrium 
at RHIC energies. The figures and curves are organized in the same way
as those in Fig. 2.

\end{itemize}

\vskip 1cm

\section*{Table 1}

\begin{center}
\begin{tabular}{|c|c|c|} \hline\hline
\multicolumn{3}{|c|}{Initial Conditions} \\ \hline
\emph{} & \emph{RHIC} & \emph{LHC} \\ \hline
$\t_0$ (fm/c)  &  0.7  &  0.5  \\ 
$\e_0$ (GeV/$\mbox{\rm fm}^3$) & 3.2   & 40.0 \\
$n_0 (\mbox{\rm fm}^{-3})$  & 2.15  & 18.0 \\
$T_0$ (GeV) & 0.50 & 0.74 \\
$l_0$  & 0.08 & 0.21 \\
$\q_0$ (fm/c) & 2.69 & 0.84 \\ \hline\hline
\end{tabular}
\end{center}

TABLE 1. Initial conditions for the evolution at RHIC and at LHC
\vskip 1cm

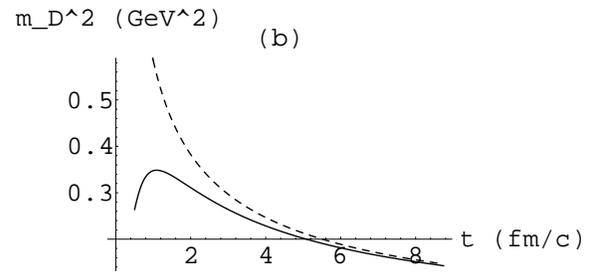
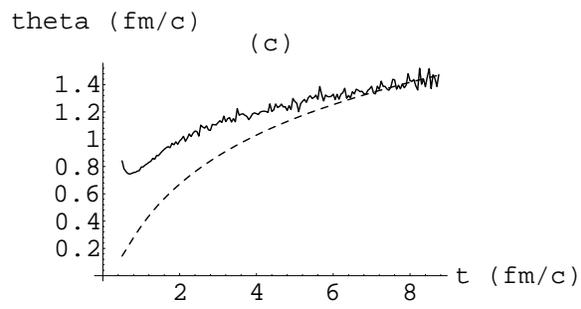
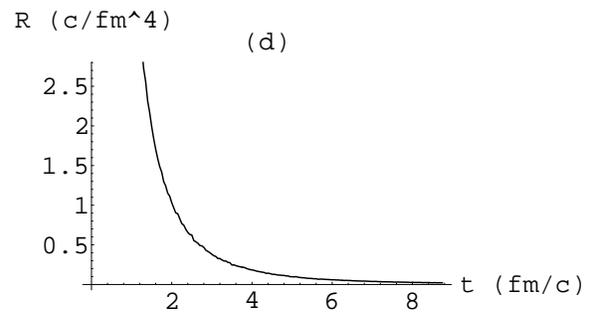
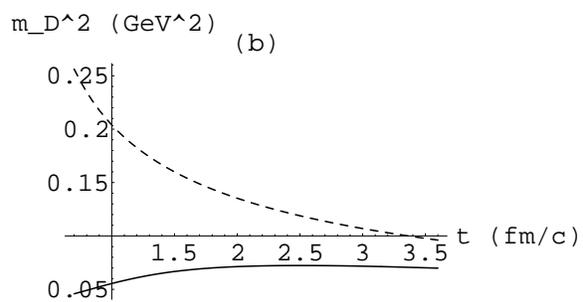
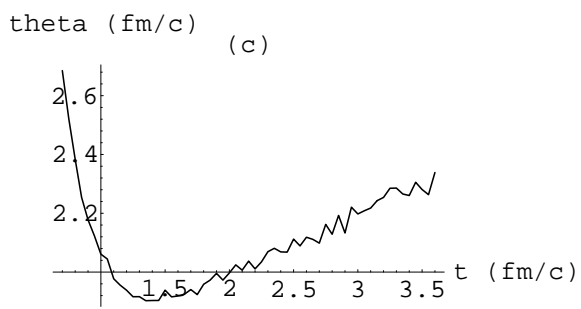
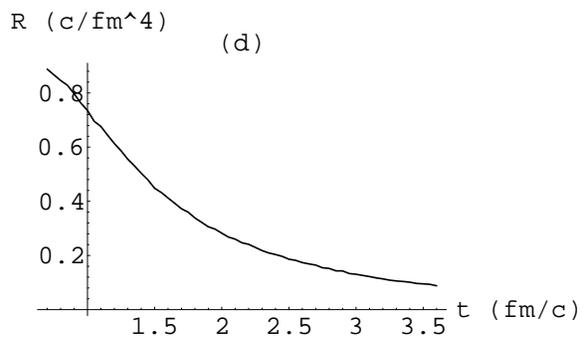
\end{document}